\documentclass[journal]{IEEEtran}
\usepackage{cite}

\newcommand\beq{\begin{equation}}
\newcommand\eeq{\end{equation}}

\usepackage{graphicx}% Include figure files
\usepackage{dcolumn}% Align table columns on decimal point
\usepackage{bm}% bold math

\usepackage{amsmath}

\usepackage{array}

\usepackage{fixltx2e}

\usepackage{stfloats}
\begin{document}
\title{Harnessing Spectral Singularities in Non-Hermitian Cylindrical Structures}

\author{Massimo~Moccia,
        Giuseppe~Castaldi,
        Andrea~Al\`u,
        and  Vincenzo Galdi
        % <-this % stops a space
\thanks{M. Moccia, G. Castaldi, and V. Galdi are with the Fields \& Waves Lab, Department of Engineering, University of Sannio, I-82100 Benevento, Italy (e-mail: massimo.moccia@unisannio.it, castaldi@unisannio.it, vgaldi@unisannio.it).}% <-this % stops a space
\thanks{A. Al\`u is with the Photonics Initiative, Advanced
	Science Research Center, the Physics Program, Graduate Center, and the
	Department of Electrical Engineering, City College, all at the City University
	of New York, New York, NY 10031, USA (email: aalu@gc.cuny.edu)}% <-this % stops a space
%\thanks{Manuscript received ; revised }
}

\date{\today}% It is always \today, today,

%%%%%%%%%%%%%%%%%%%% Created: 16/05/2018
%%%%%%%%%%%%%%%%%%%% Last revised: 26/06/2019

%\markboth{IEEE Transactions on Antennas and Propagation,~Vol.~??, No.~??, ??~2019}%
%{Moccia \MakeLowercase{\textit{et al.}}: Harnessing Spectral Singularities in Non-Hermitian Cylindrical Structures}



\maketitle

\begin{abstract}
Non-Hermitian systems characterized by suitable spatial distributions of gain and loss can exhibit ``spectral singularities'' in the form of zero-width resonances associated to real-frequency poles in the scattering operator. Here, we study this intriguing phenomenon in connection with cylindrical geometries, and explore possible applications to controlling and tailoring in unconventional ways the scattering response of sub-wavelength and wavelength-sized objects. Among the possible implications and applications, we illustrate the additional degrees of freedom available in the 
scattering-absorption-extinction tradeoff, and address the
engineering of
zero-forward-scattering, transverse scattering, and gain-controlled reconfigurability of the scattering pattern, also paying attention to stability issues.
Our results may open up new vistas in active and reconfigurable nanophotonics platforms.
\end{abstract}


% \fi
\IEEEpeerreviewmaketitle

%%%%%%%%%%%%%%%%%%%%%%%%%%%%%%%%%%%%%%%%%%%%%%%%%%%%%%%%%%%%%%
\section{Introduction}
%%%%%%%%%%%%%%%%%%%%%%%%%%%%%%%%%%%%%%%%%%%%%%%%%%%%%%%%%%%%%%

\IEEEPARstart
{T}{he control} and tailoring of the scattering response of sub-wavelength and wavelength-sized objects is a subject of pivotal interest in a variety of traditional and emerging electromagnetic scenarios, ranging from radar engineering to nanophotonics. During the last two decades, the additional degrees of freedom stemming from the increasing availability of novel materials and metamaterials \cite{Capolino:2009vr,Cai:2010om}, possibly in conjunction with the revisitation of old concepts \cite{Kerker:1975ib,Kerker:1983es}, have led to the engineering of exotic responses such as  ``invisibility cloaking'' \cite{Alu:2005at,Schurig:2006me}, ``superscattering'' \cite{Ruan:2010sl,Quian:2018ms,Quian:2019eo}, directional scattering \cite{Person:2013dz,Fu:2013dv,Yao:2016ce,Liu:2018gk,Lee:2018sn,Shamkhi:2018vo}, 
and anapoles \cite{Baryshnikova:2019oa}, just to mention a few.

It is well known that the scattering response from {\em passive} objects is subject to inherent restrictions. Some are dictated by power conservation, such as the optical theorem \cite{Kerker:1969uw,Bohren:2008wi}, which relates the forward scattering to the extinction cross-section. Moreover, passivity also dictates that the frequency poles of the scattering operator are inherently {\em complex-valued}. This implies that  scattering resonances exhibit {\em finite} lifetimes which, however, could be made in principle arbitrarily large \cite{Monticone:2014ep,Silveirinha:2014,Lannebere:2015,Rybin:2017hq,Silva:2018}.

The above restrictions are lifted for {\em active} scatterers, characterized by the presence of materials featuring gain effects (e.g., semiconductors, dyes, quantum dots). Within this framework, early studies in the 1970s have explored the possibility to attain scattering resonances characterized by real-frequency poles and zero (or negative) extinction \cite{Alexopoulos:1978es}, and addressed some potential misconceptions \cite{Kerker:1978es,Kerker:1979ri} and apparent paradoxes \cite{Kerker:1980dl,Ross:1980dl}. More recently, the interest in structures mixing active and passive constituents has been revamped under the emerging umbrella of  non-Hermitian optics \cite{Feng:2017ww} which, inspired by quantum symmetries \cite{Bender:1998rs}, has broadened the constitutive-parameter design space to the entire complex plane, indicating new pathways in the exploitation of the delicate interplay between optical gain and loss. For instance, an approach to synthesize non-Hermitian meta-atoms with unconventional scattering responses was recently proposed in \cite{Safari:2018sf}.  
Among the distinctive concepts in non-Hermitian optics, especially fascinating and ubiquitous are the so-called  ``exceptional points'' \cite{Heiss:2012ep} and ``spectral singularities'' \cite{Mostafazadeh:2009ss}. Exceptional points are spectral degeneracies implying the coalescence of eigenvalues and eigenstates. This topic has experienced a steadily growing interest in optics and photonics \cite{Miri:2019cf}, and has also started resonating in the antennas and propagation community \cite{Othman:2017to,Hanson:2019ep}. Spectral singularities are instead real frequencies at which the scattering coefficients tend to infinity, and are thus more akin to the aforementioned early observations \cite{Alexopoulos:1978es,Kerker:1978es,Kerker:1979ri,Kerker:1980dl,Ross:1980dl} of ``zero-width''  resonances. In other words, they correspond to lasing (or, in their time-reversed version, to coherent perfect absorption \cite{Chong:201ycp}) at the threshold gain, and can be observed in diverse non-Hermitian scenarios including waveguides \cite{Mostafazadeh:2009ss,Li:2019}, semi-infinite lattices \cite{Longhi:2009ss}, cylindrical \cite{Mostafazadeh:2013ss} and spherical \cite{Mostafazadeh:2011ss} scatterers, possibly in conjunction with nonlinear \cite{Mostafazadeh:2013ns}, unidirectional \cite{Ramezani:2014us},  and nonreciprocal \cite{Ramezani:2014ul} effects. Although the terminology is not always consistent, and exceptional points are sometimes defined as spectral singularities, we remark that, according to the definitions above, they are distinct concepts. One key difference is that exceptional points can also occur in purely lossy scenarios, whereas spectral singularities can only occur in the presence of gain. The reader is also referred to \cite{Mostafazadeh:2014po} for a comprehensive discussion of the mathematical foundations and physical implications of spectral singularities, and to \cite{Li:2019} for an example of  a system which can exhibit either exceptional points or spectral singularities for different parameter configurations.

In this paper, we revisit the concept of spectral singularities in connection with non-Hermitian cylindrical geometries, and illustrate how their unique properties can be harnessed in order to control and tailor in a broad fashion the scattering response. To this aim, after outlining the problem in Sec. \ref{Sec:Math}, we study in Sec. \ref{Sec:SS}  a cylindrical core-shell geometry combining gain and loss. Starting from the infinite-shell limit, which is amenable to a semi-analytical modeling and provides useful insights in the underlying phenomenology, we illustrate the spectral-singularity phenomenon and its implications in the tradeoff among scattering, absorption, and extinction.
Next, in Sec. \ref{Sec:RR}, we explore possible applications to shaping and/or reconfiguring the scattering pattern, for sub-wavelength and wavelength-sized objects, also considering stability and feasibility issues. Finally, in Sec. \ref{Sec:Conclusions}, we provide some concluding remarks and discuss the potential perspectives of our approach.

%%%%%%%%%%%%%%%%%%%%%%%%%%%%%%%%%%%%%%%%%%%%%%%%%%%%%%%%%%%%%%
\section{Problem Geometry and Statement}
%%%%%%%%%%%%%%%%%%%%%%%%%%%%%%%%%%%%%%%%%%%%%%%%%%%%%%%%%%%%%%
\label{Sec:Math}

%############################################################
%                Figure1
%
\begin{figure}
	\centering
	\includegraphics[width=\linewidth]{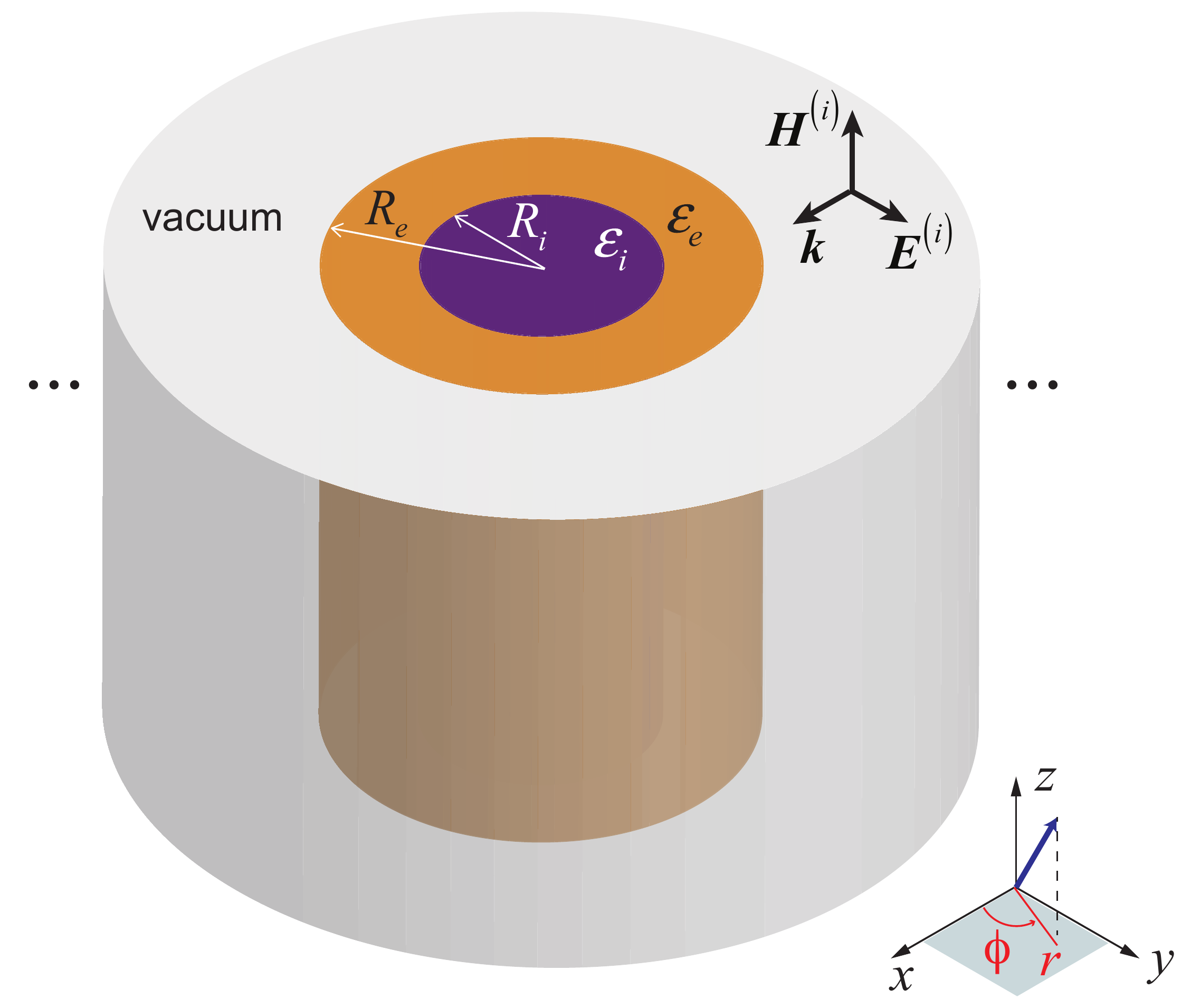}
	\caption{Problem geometry: A cylindrical core-shell structure alternating  gain and loss regions in a vacuum background, illuminated by a plane wave. Also shown are the Cartesian and associated cylindrical reference systems.}
	\label{Figure1}
\end{figure}
%############################################################

%-------------------------------------------------------------
\subsection{Geometry}
%-------------------------------------------------------------
Referring to Fig. \ref{Figure1} for schematic illustration, we consider a non-Hermitian cylindrical structure consisting of a concentric core-shell geometry with internal and external radii $R_i$ and $R_e$, respectively, of infinite extent along the $z$-direction, and immersed in vacuum. The materials in the core and shell regions are assumed as homogeneous, isotropic, nonmagnetic and characterized by complex-valued relative permittivities $\varepsilon_i=\varepsilon^\prime_i+i\varepsilon^{\prime\prime}_i$ and $\varepsilon_e=\varepsilon^\prime_e+i\varepsilon^{\prime\prime}_e$, with the prime and double-prime symbols denoting the real and imaginary part, respectively.
For the implied time-harmonic dependence $\exp(-i\omega t)$, positive and negative values of the imaginary parts of the permittivities correspond to loss and gain, respectively. In what follows, we are mainly concerned with configurations mixing loss and gain, i.e., $\varepsilon^{\prime\prime}_i>0$ and $\varepsilon^{\prime\prime}_e<0$ or vice versa. Although some previous studies on spectral singularities have assumed parity-time-symmetric scenarios \cite{Mostafazadeh:2009ss}, this is not a necessary condition and will not be specifically considered here.
Moreover, to exclude from our study the well-known case of surface-plasmon resonances (SPRs) \cite{Tribelsky:2006al}, we also assume {\em positive} real-parts of the permittivities ($\varepsilon^{\prime}_{i,e}>0$). 
In previous studies \cite{Savoia:2014to,Savoia:2015pt,Savoia:2016nh}, we have shown that planar interfaces in such scenarios  can sustain exponentially bound surface waves similar to surface plasmon polaritons. The geometry in Fig. \ref{Figure1} can be viewed as the cylindrically wrapped version of those planar scenarios.

%-------------------------------------------------------------
\subsection{Statement}
%-------------------------------------------------------------
We assume a unit-amplitude plane-wave illumination, impinging along the positive $x$-direction, with transverse magnetic (TM) polarization,  characterized by a $z$-directed magnetic field
\beq
H_z^{(i)}\!\left(r,\phi\right)=\exp\left(i k x\right)=
\sum_{m=-\infty}^{\infty}
i^m J_m\left(k r\right)\exp\!\left(im\phi\right),
\label{eq:incf}
\eeq
where $k=\omega/c=2\pi/\lambda$ denotes the vacuum wavenumber (with $c$ and $\lambda$ indicating the corresponding wavespeed and wavelength). Moreover, in view of the problem symmetry, a canonical cylindrical-wave expansion \cite{Kerker:1969uw} is applied, with $J_m$ denoting the $m$th-order Bessel functions \cite[Chap. 9]{Abramowitz:1965ao}.

We are interested in studying the scattering problem above for non-Hermitian core-shell geometries featuring both loss and gain, and in exploring the possible emergence of spectral singularities \cite{Mostafazadeh:2009rp}. This translates into determining the conditions under which the system may exhibit zero-width resonances, i.e., the scattering coefficients may admit poles on the real frequency axis.

%%%%%%%%%%%%%%%%%%%%%%%%%%%%%%%%%%%%%%%%%%%%%%%%%%%%%%%%%%%%%%
\section{Spectral Singularities}
%%%%%%%%%%%%%%%%%%%%%%%%%%%%%%%%%%%%%%%%%%%%%%%%%%%%%%%%%%%%%%
\label{Sec:SS}

%-------------------------------------------------------------
\subsection{Mathematical Formalism}
%-------------------------------------------------------------
The plane-wave scattering from radially stratified cylindrical structures is a well-known canonical problem, which can be solved analytically by using the Mie formalism \cite{Bussey:1975sb}. We begin by expanding the scattered magnetic field as
\beq
H^{(s)}_z\!\left(r,\phi\right)=\sum_{m=-\infty}^{\infty} h_{zm}\!\left(r\right) \exp\!\left(im\phi\right), 
\eeq
where the radial wavefunctions are given by
\beq
h_{zm}\!\left(r\right)=
\!\left\{
\begin{array}{lll}
	a_m J_m\!\left(k_ir\right),\quad\quad\quad\quad\quad\quad\quad\quad~~ r<R_i,\\
	b_m^{(1)} H_m^{(1)}\!\left(k_er\right)+
	b_m^{(2)} H_m^{(2)}\!\left(k_er\right)\!,~~R_i\!<r\!<R_e,\\
	i^m c_m H_m^{(1)}\!\left(kr\right), \quad\quad\quad\quad\quad\quad~~~~ r>R_e.
\end{array}
\right.
\label{eq:radwf}
\eeq
In (\ref{eq:radwf}), $k_i=k\sqrt{\varepsilon_i}$ and $k_e=k\sqrt{\varepsilon_e}$ are the wavenumbers in the core and shell regions, respectively, $H_m^{(1)}$ and $H_m^{(2)}$ denote the $m$th-order Hankel functions of first and second kind \cite[Chap. 9]{Abramowitz:1965ao}, respectively, and $a_m$, $b_m^{(1)} $, $b_m^{(2)} $ and $c_m$ are unknown expansion coefficients. These latter are determined by enforcing the continuity of the tangential components of the total (i.e., incident + scattered) electric and magnetic fields at the interfaces $r=R_i$ and $r=R_e$. Computationally convenient recursive procedures are also available \cite{Bussey:1975sb}.

%-------------------------------------------------------------
\subsection{Infinite-Shell Limit}
%-------------------------------------------------------------
It is instructive to start the study by considering the case of a cylinder embedded in a homogeneous background, which is amenable to a physically insightful analytic approximation. This can be interpreted as the infinite-shell limit ($R_e\rightarrow \infty$) of the core-shell geometry in Fig. \ref{Figure1}. 
Accordingly, the corresponding solution can be derived directly from the first two equations in (\ref{eq:radwf}) by selecting the proper behavior as $r\rightarrow\infty$. To avoid dealing with a controversial branch-cut choice in an infinite region of gain material \cite{Lakhtakia:2007wd}, 
 it makes sense to assume a lossy background ($\varepsilon''_e>0$) and the cylinder made of gain material (i.e., $\varepsilon''_i<0$). In this case, it is sufficient to set $b_m^{(2)}=0$ in (\ref{eq:radwf}) to obtain the usual radiation condition and decay at infinity. Moreover, for notational compactness, we assume $R_i=R$ in the rest of this section. The scattering coefficients in the exterior region assume the canonical form \cite{Tribelsky:2006al}
\beq
b_m^{(1)}=-\frac{\alpha_m}{\alpha_m+i\beta_m},
\label{eq:bm1}
\eeq
where
\begin{subequations}
\begin{eqnarray} 
\!\!\!\!\!\alpha_{m} &\!\!\!\!\!=\!\!\!\!\!&\sqrt{\varepsilon_i} J_m\!\left(k_iR\right) \dot{J}_{m}\!\left(k_eR\right)\!-\!\sqrt{\varepsilon_e} \dot{J}_{m}\!\left(k_iR\right)J_{m}\!\left(k_eR\right)\!,\\ 
\!\!\!\!\!\beta_{m} &\!\!\!\!\!=\!\!\!\!\!&\sqrt{\varepsilon_i} J_m\!\left(k_iR\right) \dot{Y}_{m}\!\left(k_eR\right)\!-\!\sqrt{\varepsilon_e} \dot{J}_{m}\!\left(k_iR\right)Y_{m}\!\left(k_eR\right)\!,
\end{eqnarray}
\label{eq:alphabetam}
\end{subequations}
with the overdot denoting differentiation with respect to the argument.

It is well known that, for {\em passive} scenarios, the denominator in (\ref{eq:bm1}) cannot generally vanish for real frequencies, and hence the scattering poles must be complex-valued.
Conventional Mie resonances occur for $\beta_m=0$, which results in real-valued and unit-amplitude scattering coefficients \cite{Tribelsky:2006al}.
 Notable exceptions are lossless ($\varepsilon''_{i,e,}=0$) ``plasmonic voids'' \cite{Alu:2010cw}  characterized by $\varepsilon_i>0$ and $\varepsilon_e<0$, for which the denominator in (\ref{eq:bm1}) is purely imaginary and can vanish for real frequencies.
 
 %############################################################
 %                Figure2
 %
 \begin{figure*}
 	\centering
 	\includegraphics[width=.8\linewidth]{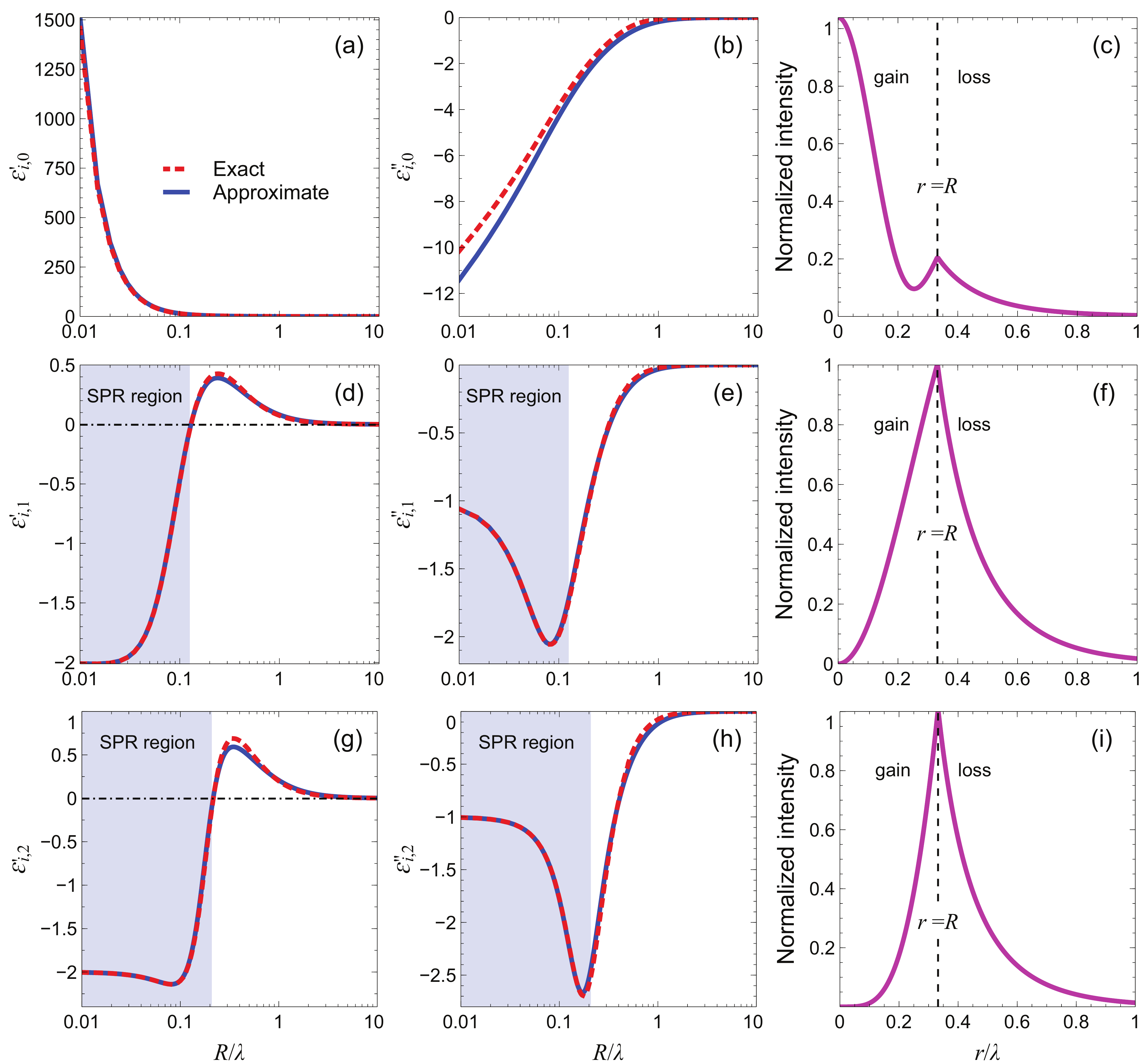}
 	\caption{Representative results for the infinite-shell case. (a), (b) Exact (red-dashed) and approximate [from (\ref{eq:epsi1}); blue-solid] real and imaginary part, respectively, of the interior relative permittivity $\varepsilon_i$, as a function of the electrical radius $R/\lambda$, for the multipolar order $m=0$ and an exterior medium with relative permittivity $\varepsilon_e=2+i$. (c) Normalized intensity of the corresponding radial wavefunction [from (\ref{eq:radwf})] for $R=0.332\lambda$ and $\varepsilon_{i,0}=2-i$. 
 	(d)--(f) and (g)--(i) Same as in panels (a)--(c), but for the multipolar orders $m=1$ and $m=2$, respectively. 
The interior relative permittivity values for the radial wavefunctions in panels (f) and (i) are $\varepsilon_{i,1}=0.374-i0.402$ and $\varepsilon_{i,2}=0.683-i1.253$, respectively. The shaded areas in panels (d), (e), (g) and (h) delimit the SPR regions characterized by negative values of the permittivity real part. The radial wavefunctions in panels (c), (f), (i) are normalized so as to remove the singular behavior at the spectral singularity; the specific normalization factors are $a_0$ for $m=0$, and $a_mJ_m\left(k_iR\right)$ for $m=1,2$.}
 	\label{Figure2}
 \end{figure*}
 %############################################################
 
 As anticipated, for the {\em non-Hermitian} gain-loss scenario of interest here, the scattering coefficients in (\ref{eq:bm1}) can instead exhibit real-frequency poles (spectral singularities). Within this framework, it is expedient to recast the pole condition
 \beq
 \alpha_m+i\beta_m=0
 \eeq
 by substituting (\ref{eq:alphabetam}) and dividing by $J_m(k_iR) H_m^{(1)}(k_eR)$ (which does not vanish for the implied branch-cut choices \cite{Abramowitz:1965ao,Doring:1966cz}). This yields a compact form
\beq
\sqrt{\varepsilon_e} F_m\left(k_iR\right)-
\sqrt{\varepsilon_i} G_m\left(k_eR\right)=0,
\label{eq:SS1}
\eeq
which contains the logarithmic derivatives
\begin{subequations}
\begin{eqnarray}
F_\nu\left(\xi\right)&=&\frac{d}{d\xi}\log\left[J_\nu\!\left(\xi\right)\right]=
\frac
{\dot{J}_\nu\!\left(\xi\right)}
{J_\nu\!\left(\xi\right)},\\
\label{eq:Fn}
G_\nu\left(\xi\right)&=&\frac{d}{d\xi}\log\left[H_\nu^{(1)}\!\left(\xi\right)\right]=
\frac
{\dot{H}_\nu^{(1)}\!\left(\xi\right)}
{H_\nu^{(1)}\!\left(\xi\right)},
\label{eq:Gn}
\end{eqnarray}
\end{subequations}
and can be solved analytically in an approximate fashion. As detailed in Appendix \ref{Sec:AppA}, in the small-argument limit $|k_iR|\ll1$, by substituting in (\ref{eq:SS1}) suitable Pad\'e-type rational approximants for the Bessel logarithmic derivatives $F_m$, we can readily solve for the interior relative permittivity $\varepsilon_i$, viz., 
	\beq
	\varepsilon_{i,m}\approx\left\{
	\begin{array}{lll}
		\displaystyle{\frac{24\left[2G_0\left(k_eR\right)+k_eR\right]}{k^2R^2 \left[8G_0\left(k_eR\right)+k_eR\right]}},\quad m=0,\\
		~\\
		\displaystyle{\frac{2m\left(m+1\right)\varepsilon_e}
		{k_eR\left[
			2\left(m+1\right)G_m\left(k_eR\right)+k_eR
			\right]}},\quad m\ge1,
		\end{array}
	\right.
	\label{eq:epsi1}
	\eeq
where the subscript $m$ tags the different multipolar orders. Here and henceforth, due to the inherent symmetry, only $m\ge0$ values are considered. 
The expressions in (\ref{eq:epsi1}) provide simple and insightful approximations from which it is possible to set the exterior medium properties ($\varepsilon_e$), an arbitrary real frequency (i.e., real-valued $k$) and radius $R$, and directly obtain (for any multipolar order $m$) the permittivity of the interior medium that should be paired in order to attain a spectral singularity.

Figure \ref{Figure2} illustrates the behavior of the solutions above for an exterior medium with relative permittivity $\varepsilon_e=2+i$. Specifically, Figs. \ref{Figure2}a and \ref{Figure2}b show the real and imaginary parts, respectively, of the interior relative permittivity pertaining to the $m=0$ order, as a function of the electrical radius. The approximate solution in (\ref{eq:epsi1}) agrees fairly well with the exact numerical solution (obtained via the \texttt{FindRoot} routine in Mathematica \cite{Mathematica}) up to rather large values of the electrical radius $R\sim10\lambda$; this is not inconsistent with our underlying assumption $|k_iR|\ll1$, as it is evident from Figs. \ref{Figure2}a and \ref{Figure2}b that the interior permittivity asymptotically vanishes for $R/\lambda\gg1$. As it can be expected, the interior medium is characterized by gain ($\varepsilon''_{i,0}<0$). More interestingly, the permittivity real part is always positive ($\varepsilon'_{i,0}>0$), thereby indicating that this phenomenon differs fundamentally from plasmonic resonances, and is essentially sustained by the imaginary-part contrast. To make this even more apparent, we can determine from Fig. \ref{Figure2}a a specific value of the electrical radius $R=0.332\lambda$ for which the real parts of the interior and exterior permittivities are {\em identical} ($\varepsilon'_i=\varepsilon'_e=2$), and therefore the resonance is solely sustained by the imaginary-part contrast. The corresponding radial wavefunction, shown in Fig. \ref{Figure2}c, is peaked at $r=0$ and decays exponentially in the exterior medium. It is also worth highlighting that the interior permittivity diverges in the small-radius limit (see the discussion below for more details), thereby indicating that the $m=0$ spectral singularity is inherently a {\em volume-type} resonance.

Figures \ref{Figure2}d--\ref{Figure2}f and \ref{Figure2}g--\ref{Figure2}i show the corresponding results for the multipolar orders $m=1$ and $m=2$, respectively. Also in these cases, the agreement with the exact numerical solution is very good, and we observe that the interior medium is always characterized by gain ($\varepsilon''_i<0$). However, different from the $m=0$ case, these do not appear to be volume-type resonances. In fact, the permittivity real-part now undergoes a sign change, becoming negative below a critical value of the electrical radius. Such negative-permittivity regions correspond to rather trivial extensions of conventional SPRs, with the gain compensating for the radiation and dissipation in the exterior medium, and are therefore not of interest in our study.
Nevertheless, it is interesting to note that the SPR transition occurs around subwavelength values of the radius, and that the imaginary parts (gain) are peaked nearby the transition, and then gradually tend to vanish in the positive-permittivity region of interest. Moreover, in this region, the radial wavefunctions (see, e.g., Figs. \ref{Figure2}f and \ref{Figure2}i) are strongly peaked at the cylinder interface, thereby resembling typical plasmonic modes, although both materials exhibit positive permittivity real-parts.

%############################################################
%                Figure3
%
\begin{figure}
	\centering
	\includegraphics[width=\linewidth]{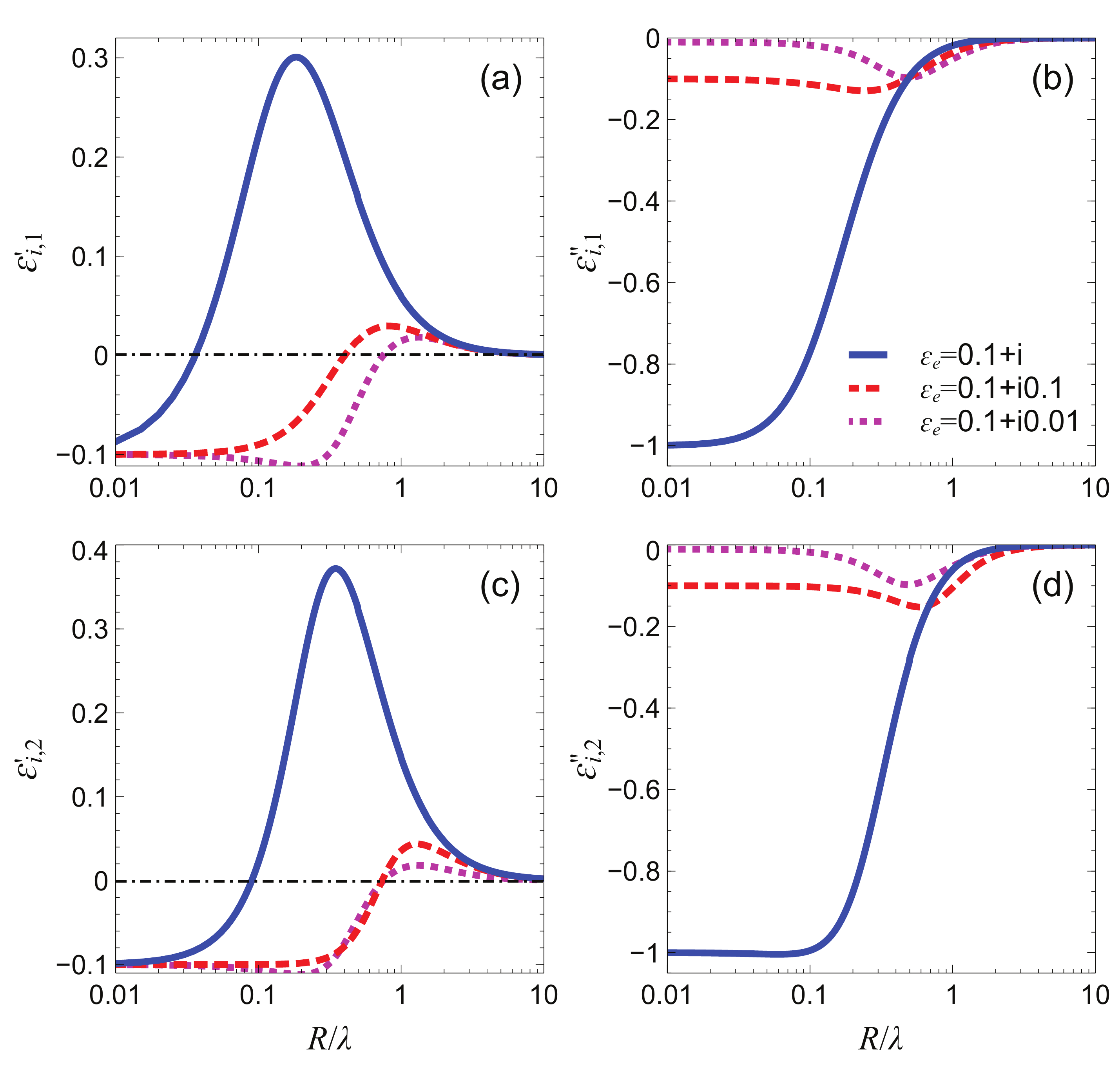}
	\caption{Representative results for the infinite-shell epsilon-near-zero case. (a), (b) Exact real and imaginary part, respectively, of the interior relative permittivity $\varepsilon_i$, as a function of the electrical radius $R/\lambda$, for the multipolar order $m=1$ and an exterior medium with relative permittivity $\varepsilon_e=0.1+i$ (blue-solid), $\varepsilon_e=0.1+0.1i$ (red-dashed), and $\varepsilon_e=0.1+0.01i$ (magenta-dotted). (c), (d) Same as in panels (a), (b), but for the multipolar order $m=2$.}
	\label{Figure3}
\end{figure}
%############################################################

It is instructive to look in more detail at the small-radius asymptotic behavior of the interior permittivities. From  (\ref{eq:epsi1}), by further substituting the small-argument approximation of the Hankel logarithmic derivatives $G_m$ (see Appendix \ref{Sec:AppA} for details), we obtain
	\beq
	\varepsilon_{i,m}\sim\left\{
	\begin{array}{lll}
		\displaystyle{\frac{6}{k^2R^2}}
		+3\varepsilon_e
		\displaystyle{\left[\log\left(\frac{-ik_eR}{2}\right)+\gamma
		\right]},\quad m=0,\\
		~\\
		\displaystyle{\frac{2m\left(m+1\right)\varepsilon_e}{k_e^2R^2-2m\left(m+1\right)}},\quad m\ge1,
	\end{array}
	\right.
	\label{eq:epsi2}
	\eeq
with $\gamma$ denoting the Euler constant.	From (\ref{eq:epsi2}), it is now apparent that, for the $m=0$ order, the divergence observed in Figs. \ref{Figure2}a and \ref{Figure2}b for the real and imaginary parts of the interior permittivity is algebraic (inverse-square) and logarithmic, respectively. Moreover, it is also evident that, for all $m\ge1$ orders, the limit $R/\lambda\rightarrow 0$ yields the SPR-type condition $\varepsilon_i=-\varepsilon_e$. Given the simple analytical structure of (\ref{eq:epsi2}), we can also readily calculate the critical radius at which the SPR transition ($\varepsilon'_i=0$) occurs, viz.,
\beq
R_{c,m}\sim \sqrt{\frac{m\left(m+1\right)\varepsilon'_e}{2}}\frac{\lambda}{\pi\left|\varepsilon_e\right|},\quad m\ge1.
\label{eq:Rc}
\eeq
While the approximation in (\ref{eq:Rc}) provides only a rough estimate, it nicely highlights the key parameters and regimes. In particular, it indicates that smaller electrical sizes can be attained by reducing the real-part of the permittivity and increasing its imaginary part, and it also implies that working in the epsilon-near-zero ($\varepsilon'_e\ll1$) regime may be beneficial, as it would substantially enhance the effects of relatively small levels of gain and losses. This is quantitatively illustrated in Fig. \ref{Figure3}, which shows the interior relative permittivity as a function of the electrical radius, for the multipolar orders $m=1$ and $m=2$, and an exterior-medium permittivity with small real part ($\varepsilon'_e=0.1$) and three representative values of the imaginary part. As it can be observed, even for small levels of gain, the SPR transition can occur around subwavelength values of the radius. 

%############################################################
%                Figure4
%
\begin{figure*}
	\centering
	\includegraphics[width=.8\linewidth]{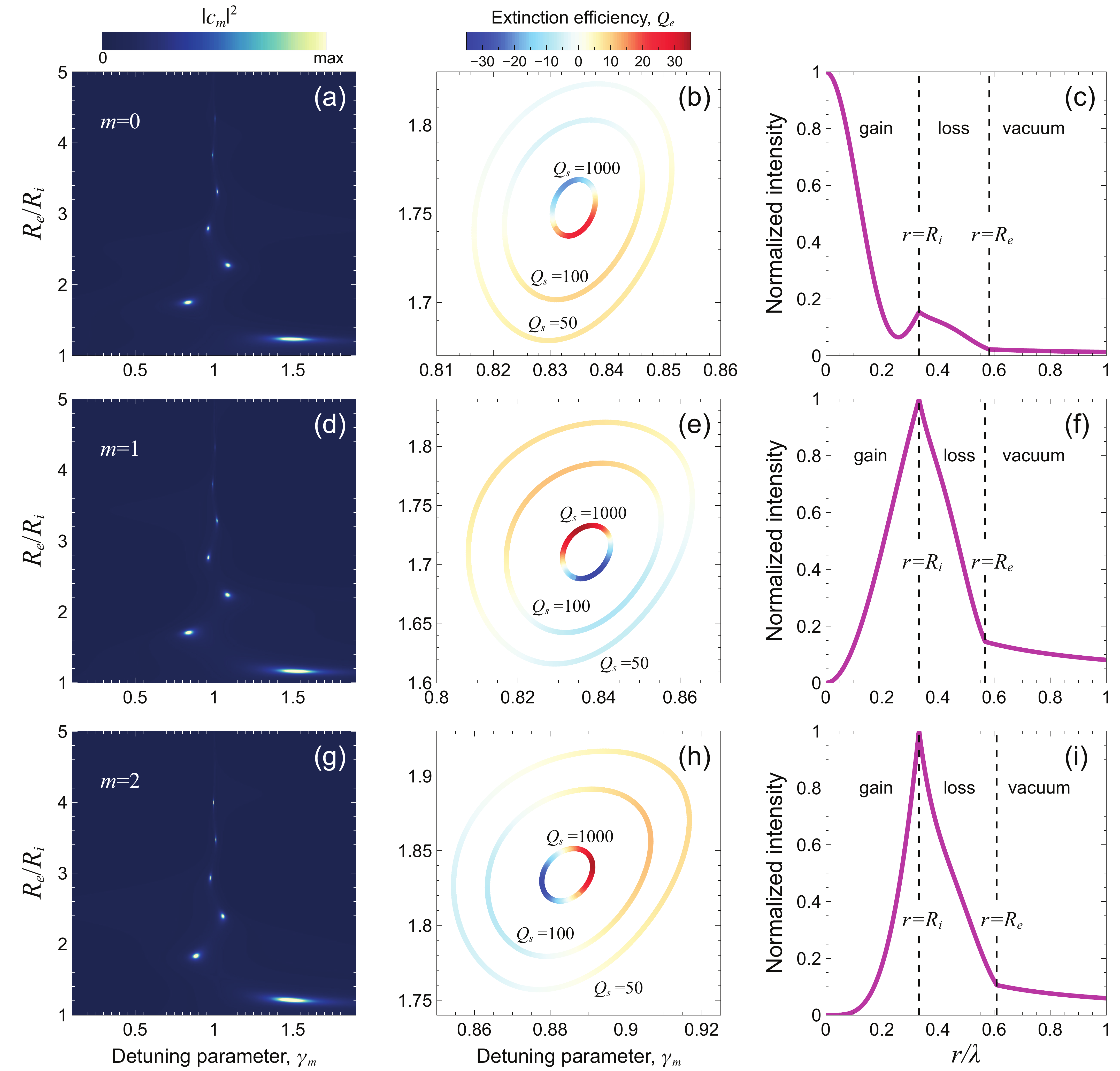}
	\caption{Representative results for a finite-shell case with interior radius $R_i=0.332\lambda$ and exterior relative permittivity $\varepsilon_e=2+i$. (a) Scattering-coefficient square-magnitude pertaining to the $m=0$ mode ($|c_0|^2$), as a function of the detuning parameter $\gamma_0$ in (\ref{eq:gammam}) and the radius ratio $R_e/R_i$, assuming $\varepsilon'_i=2$. The corresponding reference value of the interior relative permittivity is $\varepsilon_{i,0}=2-i$ (see Fig. \ref{Figure2}). (b) Iso-scattering curves, pertaining to three representative values of the scattering efficiency, displaying in false-color scale the corresponding extinction efficiency, in the vicinity of a spectral singularity. (c) Normalized intensity of a radial wavefunction [from (\ref{eq:radwf})] for $R_i=0.332\lambda$, $R_e=0.582\lambda$ (i.e., $\gamma_0=0.582$), and $\varepsilon_i=2-i0.834$. (d)--(f) As in panels (a)--(c), but for the multipolar order $m=1$ ($\varepsilon_{i,1}=0.374-i0.402$; see Fig. \ref{Figure2}). The exterior radius and interior relative permittivity values for the radial wavefunction in panel (f) are $R_e=0.568\lambda$ and $\varepsilon_i=0.374-i0.336$ (i.e., $\gamma_1=0.836$).
	(g)--(i) As in panels (a)--(c), but for the multipolar order $m=2$ ($\varepsilon_{i,2}=0.683-i1.253$; see Fig. \ref{Figure2}). The exterior radius and interior relative permittivity values for the radial wavefunction in panel (i) are $R_e=0.609\lambda$ and $\varepsilon_i=0.683-i1.108$ (i.e., $\gamma_2=0.884$).
}
	\label{Figure4}
\end{figure*}
%############################################################

Two concluding remarks are in order. First, inspired by the plasmonic analogy that relates localized surface plasmons and surface plasmon polaritons \cite{Liu:2015kk}, 
the spectral singularities above also admit an intuitive geometric interpretation as the conditions for which the surface waves that propagate at the gain-loss interface \cite{Savoia:2014to,Savoia:2015pt,Savoia:2016nh}  accumulate a phase corresponding to an integral number of $2\pi$ along the cylinder contour. Second, the dispersion equation in (\ref{eq:SS1}) in principle admits infinite branches of solutions, and the approximations in (\ref{eq:epsi1}) pertain to the lowest-order one.
Higher-order branches can be in principle approximated as well by considering higher-order rational approximants (see Appendix \ref{Sec:AppA}), although the analytical expressions would become very cumbersome. The corresponding interior permittivities and radial wavefunctions, not shown here for brevity, are characterized by increasingly larger values and faster oscillations in the cylinder region, respectively.

%-------------------------------------------------------------
\subsection{Finite Shell}
%-------------------------------------------------------------
We now move on to studying the actual core-shell configuration (with finite $R_e$) in Fig. \ref{Figure1}. In this scenario, a particularly meaningful physical observable is the bistatic scattering width \cite{Kerker:1969uw,Bohren:2008wi}
\beq
W_s\left(\phi\right)=\frac{\lambda}{2\pi}\left|
\sum_{m=-\infty}^{\infty}c_m\exp\left(im\phi\right)
\right|^2,
\eeq
with the scattering coefficients $c_m$ defined in (\ref{eq:radwf}). Other useful observables are
the total scattering, extinction and absorption widths \cite{Kerker:1969uw,Bohren:2008wi}
\begin{subequations}
	\begin{eqnarray}
	{\bar W}_s&=&\frac{\lambda}{2\pi}
	\left(
	|c_0|^2+2\sum_{m=1}^{\infty}|c_m|^2
	\right),\\
	{\bar W}_e&=&\frac{\lambda}{2\pi}\mbox{Re}
	\left(
	c_0+2\sum_{m=1}^{\infty}c_m
	\right),\\
	{\bar W}_a&=&{\bar W}_e-{\bar W}_s,
	\label{eq:Wa}
	\end{eqnarray}
	\label{eq:Wsea}
\end{subequations}
and corresponding efficiencies
\beq
Q_{\nu}=\frac{{\bar W}_{\nu}}{2R_e},\quad \nu=s,e,a.
\eeq
It follows straightforwardly from (\ref{eq:Wsea}) that, for passive scatterers, scattering and absorption widths/efficiencies are non-negative quantities.
Moreover, the well-known optical theorem \cite{Bohren:2008wi} relates the total extinction width to the forward scattering response, viz.,
\beq
{\bar W}_e=\frac{\lambda}{2\pi} \mbox{Re}\left[\Lambda\left(0\right)\right],
\eeq
with 
\beq
\Lambda\left(\phi\right)=\lim_{r\rightarrow\infty}\sqrt{\frac{\pi kr}{2}}\left\{H^{(s)}_z\!\left(r,\phi\right)\exp\left[i\left(\frac{\pi}{4}-kr\right)\right]\right\}
\eeq
denoting the scattering function.
In passive scenarios, this implies that perfect cancellation of the forward scattering can only be attained by lossless, completely invisible objects, with the so-called second Kerker condition \cite{Kerker:1975ib} holding only approximately in the quasi-static limit \cite{Alu:2010hd}. It is also generally accepted that the extinction efficiency approaches a constant value ($Q_e\rightarrow 2$) in the geometrical-optics limit of electrically large scatterers \cite{Bohren:2008wi}.

In the non-Hermitian scenarios of interest here, it is evident from (\ref{eq:Wsea}) that in the vicinity of a spectral singularity ($c_m\rightarrow\infty$) the scattering and absorption widths/efficiencies would diverge. In particular, while the scattering width/efficiency would obviously maintain a positive value, we observe from (\ref{eq:Wa}) that the absorption width/efficiency would be {\em negative} [since $\mbox{Re}(c_m)<|c_m|^2$], thereby implying amplification.

Figure \ref{Figure4} illustrates some representative results. In this case, no simple analytical parameterization can be derived, and we need to numerically find the scattering-coefficient poles. Nevertheless, it is still insightful and computationally effective to look for these solutions as perturbations of the infinite-shell case. Accordingly, we consider the same interior radius, exterior permittivity, and interior-permittivity real part as in Fig. \ref{Figure2}, and vary the exterior radius and interior-permittivity imaginary part. For this latter, we define a detuning parameter
\beq
\gamma_m=\frac{\varepsilon''_i}{\varepsilon''_{i,m}},
\label{eq:gammam}
\eeq
which parameterizes the departure from the exact infinite-shell values shown in Fig. \ref{Figure2}. Specifically, with reference to the multipolar order $m=0$, Fig. \ref{Figure4}a shows in false-color scale the scattering-coefficient square-magnitude ($|c_0|^2$) as a function of the radius ratio and detuning parameter. The bright spots that can be observed correspond to poles, i.e., spectral singularities. It is interesting to notice that, for increasing values of the radius ratio $R_e/R_i$, these spots become more and more localized and the corresponding detuning parameter approaches one, i.e., the infinite-shell limit. In the same parameter plane, Fig. \ref{Figure4}b shows three representative iso-scattering curves (corresponding to scattering efficiencies $Q_s=50, 100, 1000$) in the vicinity of one such spectral singularity, also displaying (in false-color scale) the corresponding extinction efficiency. We observe that, for a given scattering efficiency, the extinction efficiency can be tuned to assume either positive, negative or zero values; the absorption efficiency (not shown) remains instead always negative.
Figure \ref{Figure4}c shows a radial wavefunction at a spectral singularity. By comparison with the infinite-shell case (cf. Fig. \ref{Figure2}c), we observe a qualitatively similar behavior in the gain and loss regions, with the expected algebraic decay in the outermost vacuum region. 
Qualitatively similar observations hold for the $m=1$ and $m=2$ multipolar orders, whose responses are shown in Figs. \ref{Figure4}d--\ref{Figure4}f and \ref{Figure4}g--\ref{Figure4}i, respectively. 

%############################################################
%                Figure5
%
\begin{figure}
	\centering
	\includegraphics[width=\linewidth]{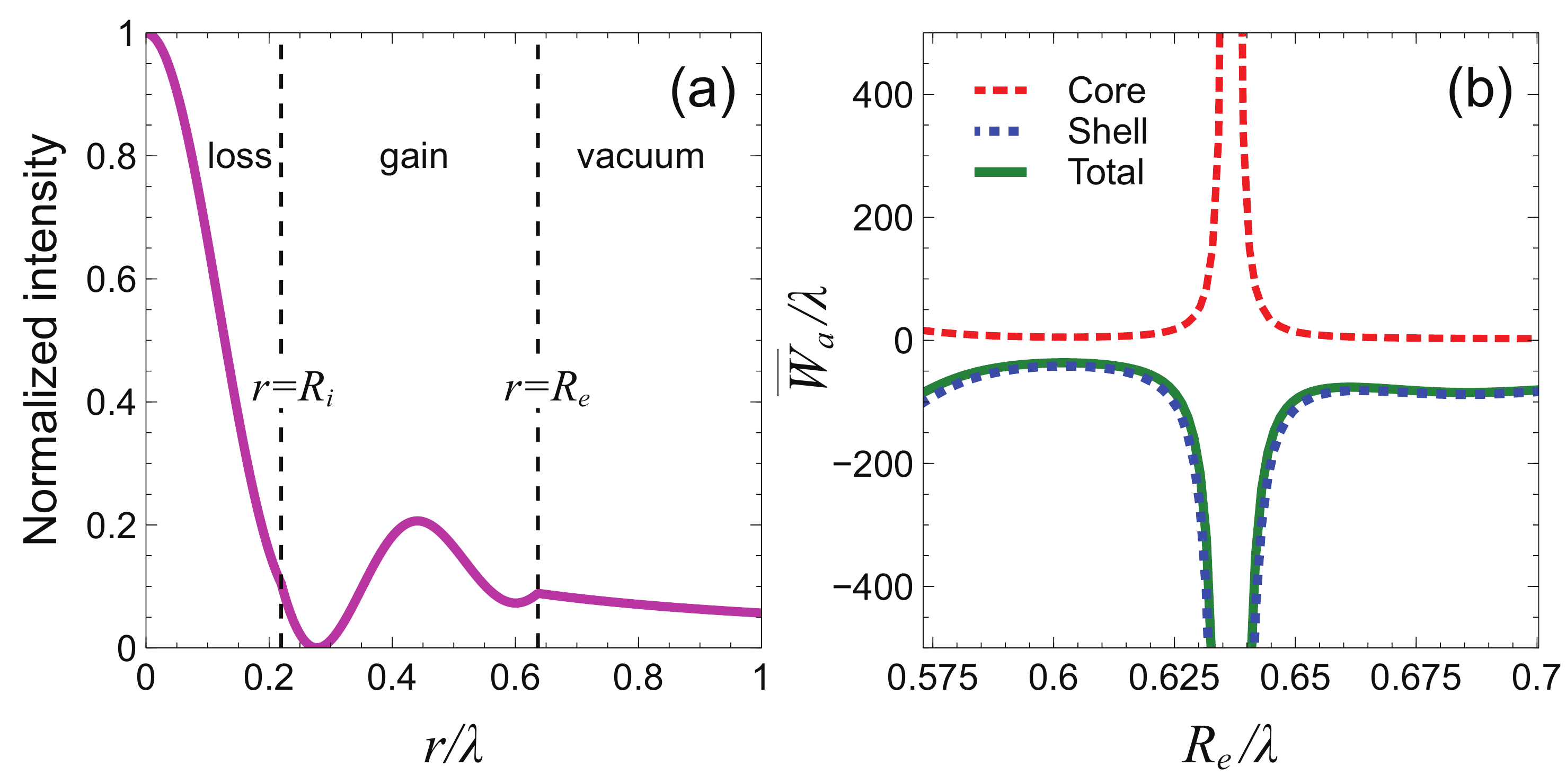}
	\caption{(a) Normalized radial wavefunction (square-magnitude) pertaining to a spectral singularity for the multipolar order $m=0$, for an inverted core-shell configuration with 
		$\varepsilon_i=2+i$, $\varepsilon_e=2-i$,
		$R_i=0.220\lambda$, and $R_e=0.637\lambda$. (b) Normalized absorption width pertaining to the core (red-dashed), shell (blue-dotted), and total (purple-solid) regions, as a function of the exterior radius; note the divergence at the spectral singularity $R_e=0.637\lambda$.}
	\label{Figure5}
\end{figure}
%############################################################

In the finite-shell scenario, we are no longer restricted to configurations with gain in the interior region, and spectral singularities can be observed also for inverted configurations with gain in the exterior region ($\varepsilon''_i>0$, $\varepsilon''_e<0$). Figure \ref{Figure5} shows a representative example pertaining to the multipolar order $m=0$ where, by comparison with Figs. \ref{Figure4}a--\ref{Figure4}c, the interior and exterior permittivities are flipped, and the radii are adjusted so as to attain a spectral singularity. From the radial wavefunction in Fig. \ref{Figure5}a, we observe a qualitatively similar behavior as in Fig. \ref{Figure4}a, with some oscillations in the shell region. Figure \ref{Figure5}b shows the total absorption width [see (\ref{eq:Wa})] as a function of the exterior radius in the vicinity of the spectral singularity. As anticipated, we observe a divergence with negative sign, which is indicative of amplification. Also shown, as references, are the two separate contributions from the core (absorption) and shell (amplification) regions, which can be computed via the Poynting theorem by directly integrating the active powers. Interestingly, they both diverge, but the shell contribution diverges faster, thereby dictating the negative sign and the overall amplification.
This implies that spectral singularities could be exploited to induce strong power dissipation (orders of magnitude larger than normal) within subwavelength lossy regions, with potentially interesting applications to enhancing nonlinear or photothermal effects.

%############################################################
%                Figure6
%
\begin{figure}
	\centering
	\includegraphics[width=\linewidth]{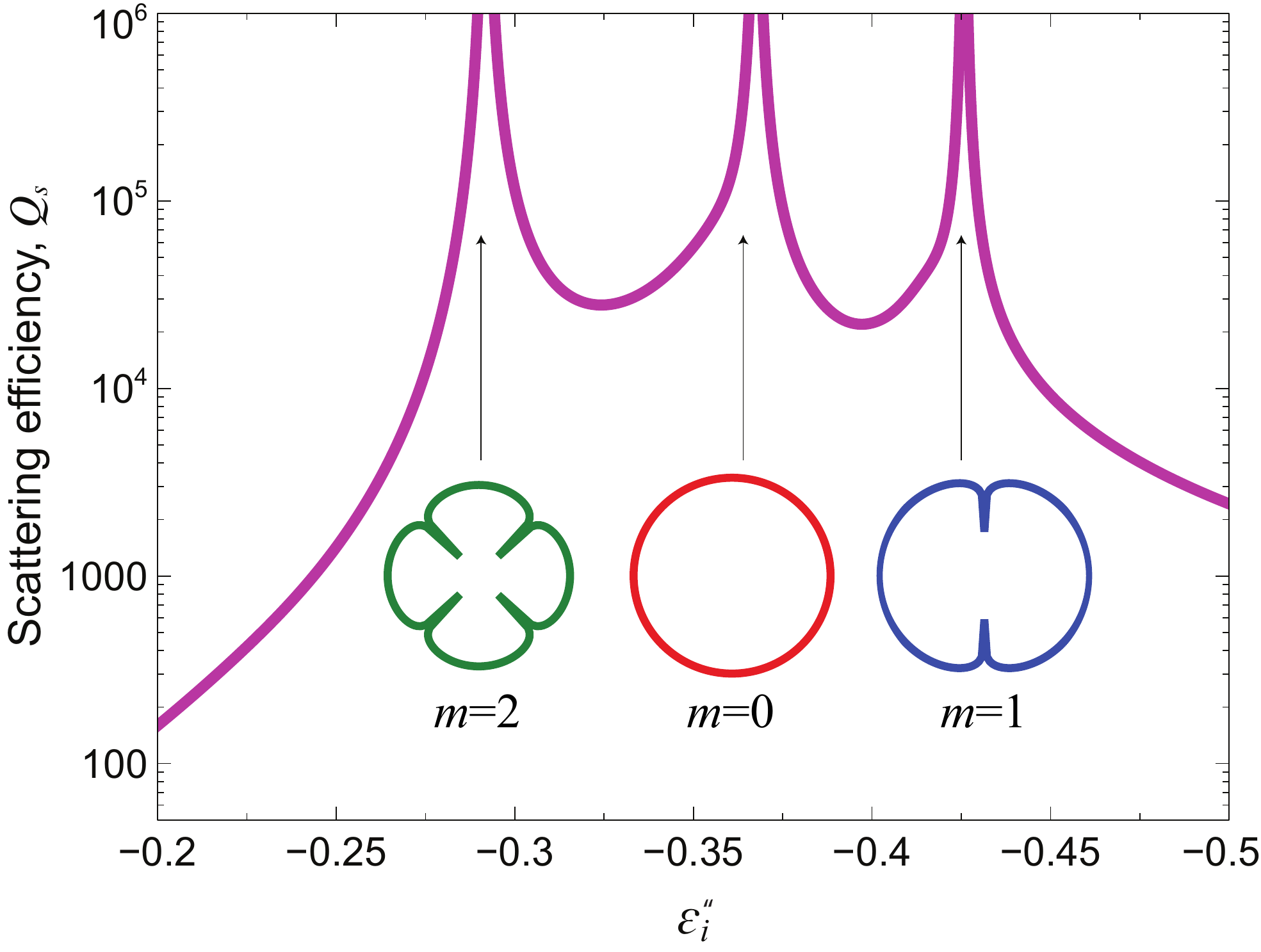}
	\caption{Scattering efficiency for a core-shell configuration with $R_i=2.646\lambda$, $R_e=2.722\lambda$, $\varepsilon_e=1.534 +i0.230$, and $\varepsilon'_i=1.460$, as a function of $\varepsilon''_i$ (gain). Note the three spectral singularities pertaining to the multipolar orders $m=0$ ($\varepsilon''_i=-0.368$), $m=1$ ($\varepsilon''_i=-0.426$), and $m=2$ ($\varepsilon''_i=-0.292$), with the corresponding scattering patterns shown as insets.}
	\label{Figure6}
\end{figure}
%############################################################

%%%%%%%%%%%%%%%%%%%%%%%%%%%%%%%%%%%%%%%%%%%%%%%%%%%%%%%%%%%%%%
\section{Applications to Pattern Shaping and Reconfigurability}
%%%%%%%%%%%%%%%%%%%%%%%%%%%%%%%%%%%%%%%%%%%%%%%%%%%%%%%%%%%%%%
\label{Sec:RR}

%-------------------------------------------------------------
\subsection{Physical Models of Gain and Loss Media}
%-------------------------------------------------------------
In what follows, for the gain media we assume the approximate linearized model derived in \cite{Campione:2011cm} for materials made of fluorescent dye molecules (modeled as four level atomic systems), viz.,
\beq
\varepsilon_G\!\left(\omega\right)\!=\!\varepsilon_{h}\!+\!\frac{\sigma_a}{\left(\omega^{2}\!+\!i \Delta\omega_a \omega\!-\!\omega_a^{2}\right)} \frac{\left(\tau_{21}-\tau_{10}\right) \Gamma_{p}\overline{N}_{0}}{\left[1\!+\!\left(\tau_{32}\!+\!\tau_{21}\!+\!\tau_{10}\right) \Gamma_{p}\right]} \!,
\label{eq:gain}
\eeq
where $\omega_a$ is the emitting radian frequency, $\Delta\omega_a$ is the bandwidth of the dye transition, $\varepsilon_h$ is the host-medium relative permittivity, $\sigma_a$ is a coupling strength parameter, $\bar{N}_0$ is the total dye concentration, $\tau_{jl}$ are relaxation times  for the state transitions, and $\Gamma_p$ is the pumping rate.

For the lossy media, we assume instead a standard Lorentz-type dispersion model:
\beq
\varepsilon_L\left(\omega\right) = {\varepsilon_\infty } - \frac{{{\omega^2_p}}}{{\left( {{\omega^2} - \omega_0^2 + i\omega\Gamma } \right)}},
\label{eq:loss}
\eeq
where $\varepsilon_\infty$ denotes the high-frequency limit, $\omega_0$ the center radian frequency, and $\Gamma$ a dampening factor.

The syntheses below, derived through the time-harmonic wave-scattering formalism in Sec. \ref{Sec:SS}, pertain to the steady-state response and consider only the permittivity values at the design frequency. However, we will consider the dispersive models in (\ref{eq:gain}) and (\ref{eq:loss}) when assessing the stability (Sec. \ref{Sec:Stability}).

%############################################################
%                Figure7
%
\begin{figure}
	\centering
	\includegraphics[width=\linewidth]{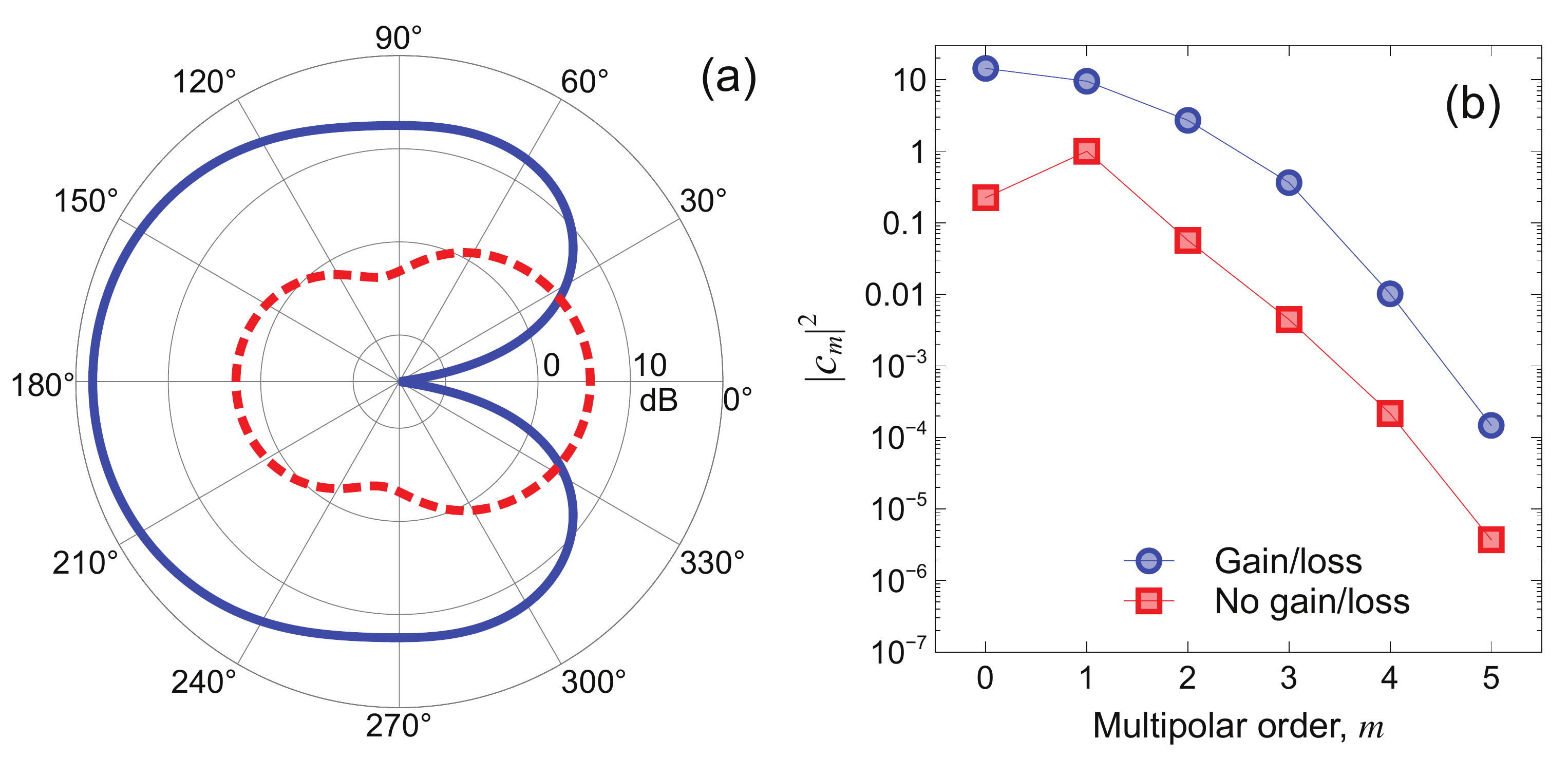}
	\caption{Example of zero-forward-scattering pattern synthesis. (a) Polar plot of bistatic scattering width (normalized with respect to the wavelength, in dB scale) pertaining to a core-shell configuration with 
		$R_i=0.164\lambda$,
		$R_e=0.482\lambda$, $\varepsilon_i=11.305+i2.741$,
		$\varepsilon_e=1.125-i0.801$ (blue-solid curve). The value at forward-scattering is $-57$dB. Also shown (red-dashed), as a reference, is the response in the absence of gain and loss (i.e., $\varepsilon_i=11.305$, $\varepsilon_e=1.125$). (b) Corresponding scattering coefficients (square magnitude) for the first six multipolar orders, in the presence (blue circles) and absence (red squares) of gain and loss, respectively. Continuous curves are guides to the eye only.}
	\label{Figure7}
\end{figure}
%############################################################

%-------------------------------------------------------------
\subsection{Examples of Syntheses}
%-------------------------------------------------------------

The core-shell geometry above possesses a number of degrees of freedom which can be exploited to harness spectral singularities and engineer exotic scattering responses. Referring to 
Appendix \ref{Sec:AppB} for details on the synthesis procedures, in what follows we illustrate some representative examples.

Figure \ref{Figure6} shows the scattering efficiency pertaining to a specifically optimized parameter configuration for which slight variations of the gain in the core region [compatible with realistic values attainable from the model in (\ref{eq:gain})] selectively excite the spectral singularities pertaining to the first three multipolar orders ($m=0,1,2$), with the corresponding scattering patterns shown as insets. This may find interesting applications in reconfigurable nanophotonic scenarios, by enabling scattering/emission-pattern reconfigurability controlled via optical pumping. Moreover, it also indicates that, even for scatterers of moderately large electrical size ($R_e=2.722\lambda$ in the example), a low number of multipolar orders (possibly one) can dominate the scattering pattern.

%############################################################
%                Figure8
%
\begin{figure}
	\centering
	\includegraphics[width=\linewidth]{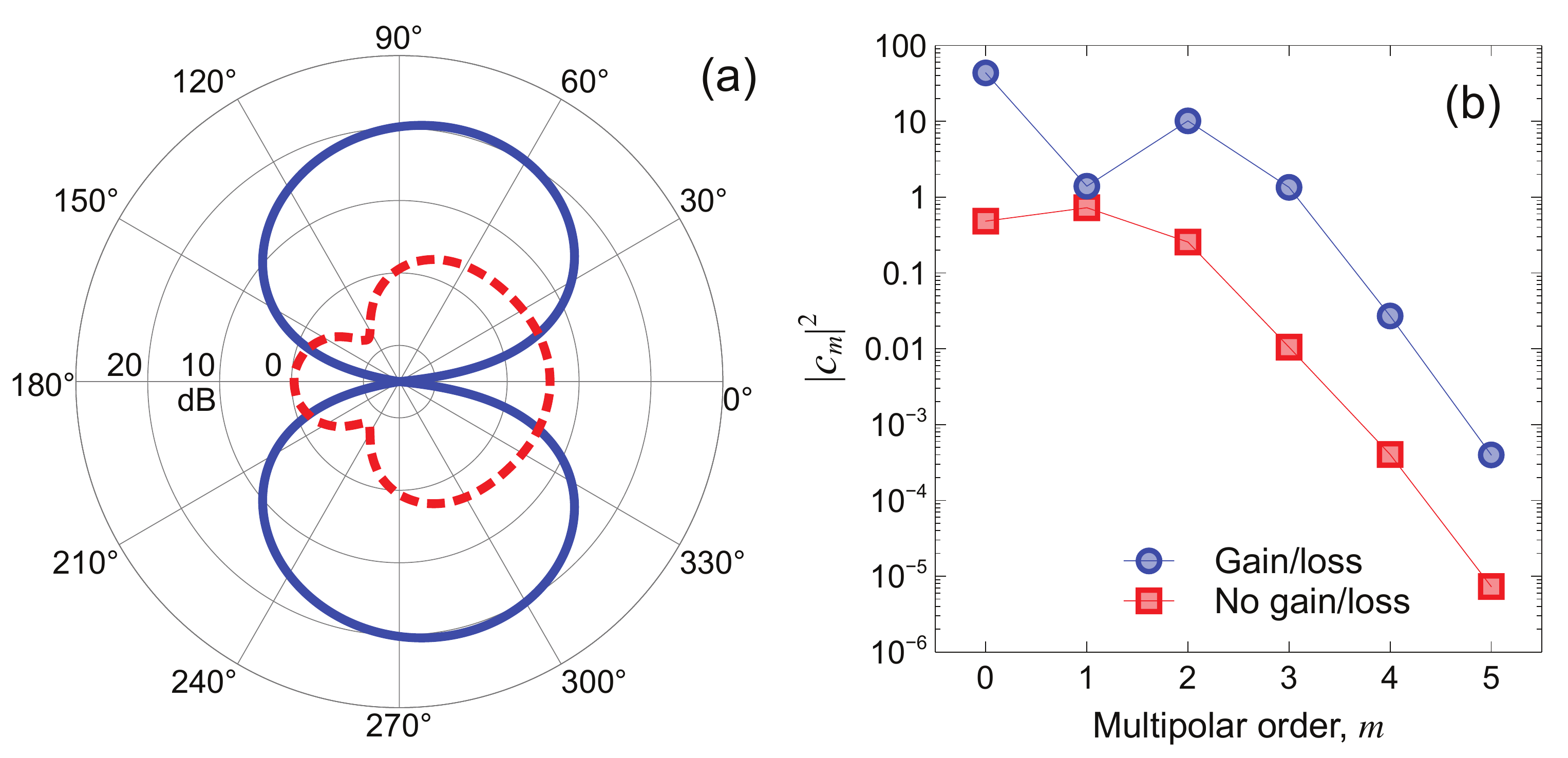}
	\caption{Example of transverse-scattering pattern synthesis. (a) Polar plot of bistatic scattering width (normalized with respect to the wavelength, in dB scale) pertaining to a core-shell configuration with 
		$R_i=0.224\lambda$,
$R_e=0.496\lambda$, $\varepsilon_i=6.993+i1.406$,
$\varepsilon_e=1.138-i1.064$ (blue-solid curve). The values at forward- and back-scattering are $-50$dB and $-58$dB, respectively. Also shown (red-dashed), as a reference, is the response in the absence of gain and loss (i.e., $\varepsilon_i=6.993$, $\varepsilon_e=1.138$). (b) Corresponding scattering coefficients (square magnitude) for the first six multipolar orders, in the presence (blue circles) and absence (red squares) of gain and loss, respectively. Continuous curves are guides to the eye only.}
	\label{Figure8}
\end{figure}
%############################################################

Clearly, working in parameter regimes very close to spectral singularities renders the system inherently prone to optical instability, i.e., to support self-oscillations. One may therefore wonder to what extent it is possible to suitably {\em detune} the spectral singularities so as to maintain the dominance of selected multipolar orders and yet do not incur in optical instabilities.
Figure \ref{Figure7} shows an interesting example in this respect, pertaining to a configuration with losses in the core and gain in the shell, whose parameters have been optimized so as to attain {\em zero-forward-scattering}. Such response, typically referred to as second Kerker condition \cite{Kerker:1983es}, can be interpreted for electrically small scatterers as the destructive interference between the electric and magnetic dipoles (Huygens source). However, for electrically larger scatterers, such interpretation is no longer valid, as higher-order multipoles are not negligible. From the synthesized scattering pattern shown in Fig. \ref{Figure7}a, we observe the shape typical of Huygens sources, even though the scatterer is wavelength-sized. Looking at the scattering coefficients in Fig. \ref{Figure7}b, we notice that the first two multipolar orders ($m=0,1$) are more strongly excited, with moderately larger value (unattainable in the passive case), and yet with the corresponding spectral singularities sufficiently detuned  in order to favor stability (see the discussion in Sec. \ref{Sec:Stability}). Thus, even for this wavelength-sized object, the scattering mechanism  is essentially dominated by the destructive interference between the two lowest multipolar orders. To better highlight the key role played by non-Hermiticity, also shown are the scattering pattern and corresponding coefficients in the absence of loss and gain, which show markedly different behaviors.

As a further example, Fig. \ref{Figure8} shows a synthesized response implementing {\em transverse} (i.e., zero forward and backward) scattering. Also in this case, in spite of a wavelength-sized object, the response is dominated by two multipolar orders ($m=0,2$), with the corresponding spectral singularities suitably detuned. Once again, by comparison with the reference response in the absence of gain and loss, the instrumental role of non-Hermiticity is highlighted.

%############################################################
%                Figure9
%
\begin{figure*}
	\centering
	\includegraphics[width=.75\linewidth]{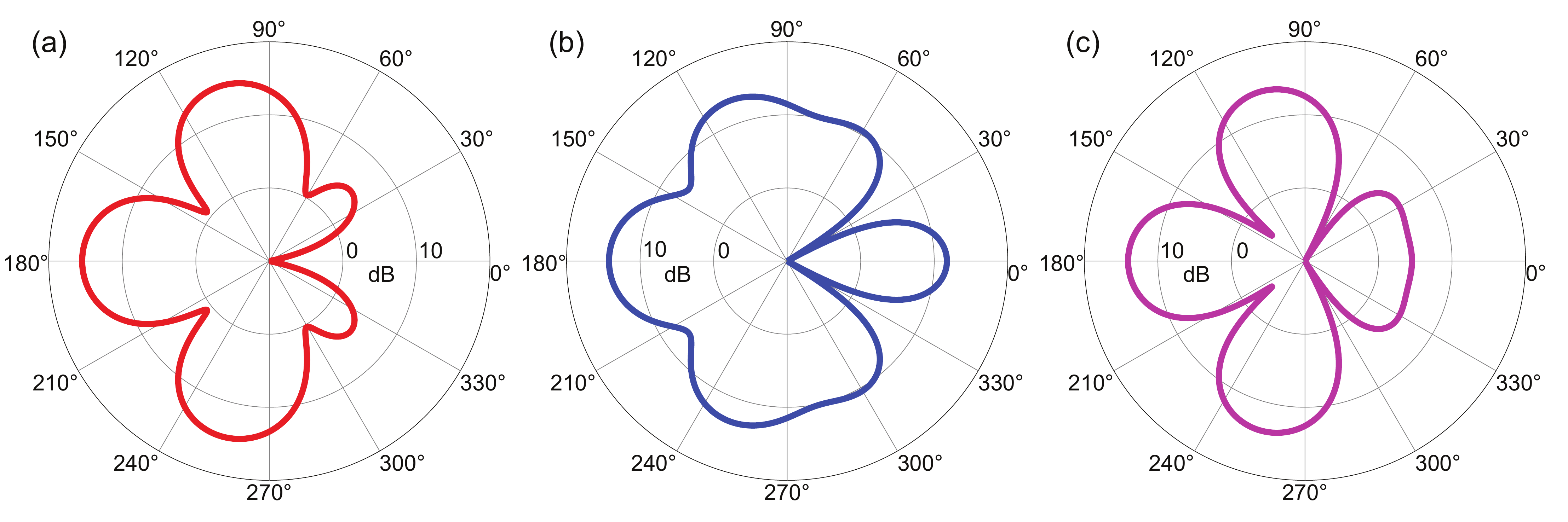}
	\caption{Example of reconfigurable scattering pattern. (a), (b), (c) Polar plots of bistatic scattering width (normalized with respect to the wavelength, in dB scale) pertaining to a core-shell configuration with 
		$R_i=0.212\lambda$,
		$R_e=0.525\lambda$, $\varepsilon_i=9.136+i2.362$, $\varepsilon'_e=1.936$, and $\varepsilon''_e=-0.850$, $-1.186$ and $-0.719$, respectively; note the scattering zeros enforced at $\phi=0$, $\phi=30^o$ and $\phi=60^o$, respectively.}
	\label{Figure9}
\end{figure*}
%############################################################

Finally, we show in Fig. \ref{Figure9} an example of scattering-pattern reconfigurability, where the position of a scattering zero is changed (from $\phi=0$, to $\phi=30^o$ or $\phi=60^o$)
by solely acting on the gain level, with all other geometrical and constitutive parameters unchanged. Also in this case the effect is obtained by selectively changing some dominant low-order multipoles (not shown for brevity) but, unlike the example in Fig. \ref{Figure6}, the spectral singularities are suitably detuned in order to favor stability (see Sec. \ref{Sec:Stability}).

%############################################################
%                Figure10
%
\begin{figure*}
	\centering
	\includegraphics[width=\linewidth]{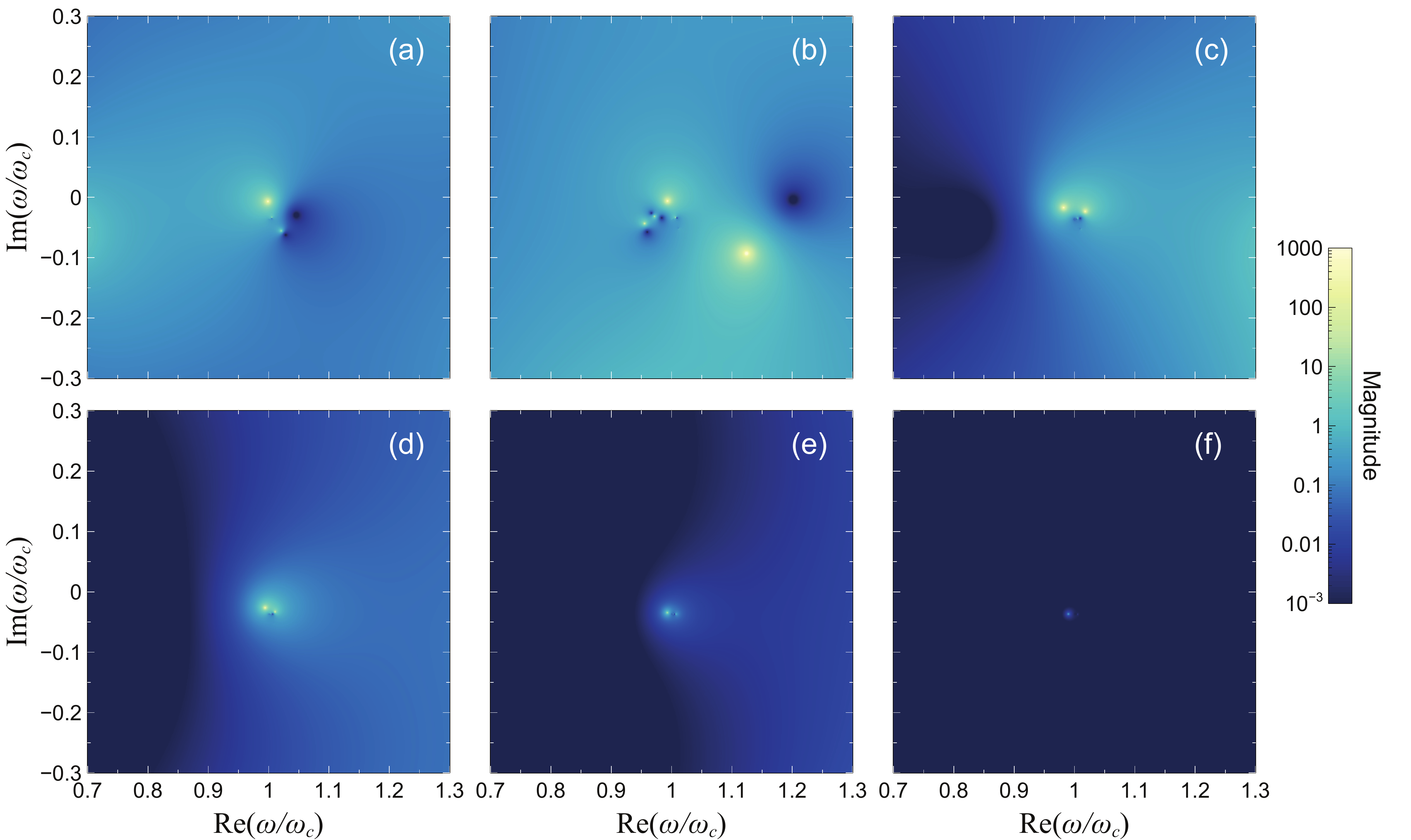}
	\caption{Stability analysis. Geometry and parameters as in Fig. \ref{Figure7}, with dispersive permittivity models as in (\ref{eq:gain}) and (\ref{eq:loss}),
		so as to satisfy the nominal design condition  the center radian frequency $\omega_c$. 
		The core (lossy) relative permittivity is obtained from (\ref{eq:loss}), with $\varepsilon_\infty=11.305$, $\omega_0=\omega_c$, $\omega_p=0.274\omega_c$ and $\Gamma=0.1\omega_c$. The shell (gain) relative permittivity is obtained from (\ref{eq:gain}), with 
		$\varepsilon_h=1.125$, $\omega_a=\omega_c=2\pi c/\lambda_a$ (with $\lambda_a=777$ nm), $\Delta \omega_a=2 \pi c \Delta\lambda_a / \lambda_a^2$ (with $\Delta\lambda_a=56$ nm), $\sigma_a=6 \pi c^{3} \eta/\left(\tau_{21} \omega_a^2 \sqrt{\varepsilon_h}\right)$ (with $\eta=0.48$),  
		$\bar{N}_0=6.1 \cdot 10^{18}$ cm$^{-3}$, $\tau_{21}=50$ ps, $\tau_{10}=100$ fs, $\tau_{32}=100$ fs, $\Gamma_p=1.536\times10^9$ s$^{-1}$.
		(a)–-(f) Magnitude (square) of scattering coefficients $|c_m|^2$, over the complex $\omega$ plane, for multipolar orders $m = 0,1,...,5$, respectively. The complex radian frequency is normalized by its center value $\omega_c$, at which the nominal design is attained. Note that the poles are all in the lower half-plane $\mbox{Im}\left(\omega\right)<0$.}
	\label{Figure10}
\end{figure*}
%############################################################

%-------------------------------------------------------------
\subsection{Stability Analysis}
%-------------------------------------------------------------
\label{Sec:Stability}
As anticipated, due to the presence of gain, our non-Hermitian configuration is prone to optical instability, which can manifest as self-oscillations supported by the system.  
In \cite{Savoia:2017mi}, for a non-Hermitian cylindrical structure composed of concentric gain/loss metasurfaces, it was shown that the system could be made {\em unconditionally stable} by suitably choosing the dispersion properties of the two metasurfaces.

Figure \ref{Figure10} shows the scattering coefficients (square magnitude) over the complex $\omega$ plane, for the $0\le m\le5$ multipolar orders relevant to the example in Fig. \ref{Figure7}, assuming the dispersive models in (\ref{eq:gain}) and (\ref{eq:loss}) with parameters (given in the caption) chosen so as to satisfy the design condition at the center radian frequency $\omega_c$. We observe that all poles are confined within the lower half of the complex plane $\mbox{Im}\left(\omega\right)<0$ which, consistently with our assumed  time-harmonic convention, implies that the system is {\em unconditionally stable} for any temporal excitation. This example only serves to prove that it is possible in principle to attain unconditional stability by suitably detuning the spectral singularities, but it is worth stressing that different  parameter choices in the dispersion models and/or the synthesis may cause the transition of poles to the upper half-plane $\mbox{Im}\left(\omega\right)>0$, thereby rendering the system unstable. Along the same lines, we have verified that unconditional stability can be attained in principle for the examples in Figs. \ref{Figure8} and \ref{Figure9}  too.

%-------------------------------------------------------------
\subsection{Technological Feasibility}
%-------------------------------------------------------------

As detailed in Appendix \ref{Sec:AppB}, although our synthesis procedure does not rely on specific material libraries, we do enforce some constraints so that the arising constitutive properties are consistent with realistic values considered in the literature. In the examples of Figs. \ref{Figure6}--\ref{Figure9}, the constitutive parameters are treated as continuous optimization variables, and therefore the outcomes do not exactly correspond to specific materials. Nevertheless, with reference to the scenario in Fig. \ref{Figure7}, the core (lossy) relative permittivity is in line with that exhibited by semiconductors such as silicon, InP, GaP at infrared wavelengths \cite{Palik:1998}, with suitable absorptive dopants. On the other hand, 
the shell (gain) relative permittivity is obtained from (\ref{eq:gain}), with parameters that are consistent with models of organic dyes \cite{Campione:2011cm,Caligiuri:2017rg} embedded in a low-index dielectric \cite{Schubert:2007,Chi:2012} available in the literature. For instance, at near-infrared wavelengths, a realistic parameter configuration could entail $\varepsilon_h=1.125$ and the
organic dye LDS798 (by Exciton) featuring $\omega_a=\omega_c=2\pi c/\lambda_a$ (with $\lambda_a=777$ nm), $\Delta \omega_a=2 \pi c \Delta\lambda_a / \lambda_a^2$ (with $\Delta\lambda_a=56$ nm), $\sigma_a=6 \pi c^{3} \eta/\left(\tau_{21} \omega_a^2 \sqrt{\varepsilon_h}\right)$ (with $\eta=0.48$),  
$\bar{N}_0=6.1 \cdot 10^{18}$ cm$^{-3}$, $\tau_{21}=50$ ps, $\tau_{10}=100$ fs, $\tau_{32}=100$ fs \cite{Caligiuri:2017rg}, and pumping rate 
$\Gamma_p=1.536\times10^9$ s$^{-1}$ compatible with the value considered in \cite{Campione:2011cm}. The gain level considered is also compatible with those attainable via quantum dots \cite{Moreels:2012,Campbell:2012}.

Clearly, the parameter configurations above only serve to illustrate that the required constitutive properties are feasible. 
With a view toward a practical implementation, a library featuring a discrete set of dispersive material models should be considered upfront, also taking into account fabrication-compatibility constraints. The resulting mixed discrete-continuous optimization problem could be addressed via hybrid (e.g., enhanced multipoint approximation)  or genetic-algorithm-based approaches \cite{Liu:2018}.

%%%%%%%%%%%%%%%%%%%%%%%%%%%%%%%%%%%%%%%%%%%%%%%%%%%%%%%%%%%%%%
\section{Conclusions and Outlook}
%%%%%%%%%%%%%%%%%%%%%%%%%%%%%%%%%%%%%%%%%%%%%%%%%%%%%%%%%%%%%%
\label{Sec:Conclusions}
To sum up, we have revisited the concept of spectral singularities in connection with non-Hermitian cylindrical structures, with a view toward harnessing their properties to tailor and control the scattering response in unconventional ways. In particular, we have illustrated the underlying physical aspects  via an insightful semi-analytical model, also highlighting the similarities and differences with respect to conventional SPR phenomena. Moreover, we have presented several possible implications and applications to the scattering-absorption-extinction tradeoff, as well as to shaping and/or reconfiguring the scattering pattern, for sub-wavelength and wavelength-sized objects. We have also shown that, via suitable parameter detuning, unconditional stability can be attained, while preserving some of the exotic traits associated with this phenomenology.

Our findings may be of interest to a variety of applications ranging from active and reconfigurable nanophotonics platforms to enhancing nonlinear or photothermal effects.

Current and future studies include the extension to different geometries (e.g., spherical) and fully 3-D (vector) scenarios, as well as the exploration of new applications.

\appendices
%%%%%%%%%%%%%%%%%%%%%%%%%%%%%%%%%%%%%%%%%%%%%%%%%%%%%%%%%%%%%%
\section{Details on Eqs. (\ref{eq:epsi1}) and (\ref{eq:epsi2})}
%%%%%%%%%%%%%%%%%%%%%%%%%%%%%%%%%%%%%%%%%%%%%%%%%%%%%%%%%%%%%%
\label{Sec:AppA}
The Bessel logarithmic derivative in (\ref{eq:Fn}) satisfies the recurrence relation \cite[Eq. (8.39)]{Bohren:2008wi} 
\beq
F_\nu\left(\xi\right)=\frac{\nu}{\xi}-\frac{1}{\displaystyle{\frac{\nu+1}{\xi}}+F_{\nu+1}\left(\xi\right)},
\eeq
which, truncated after two ($\nu=0$) and one ($\nu\ge1$) iterations, yields the Pad\'e-type rational approximants
\beq
F_{\nu}\left(\xi\right)\approx\left\{
\begin{array}{lll}
\displaystyle{\frac{\xi\left(\xi^2-24\right)}{8\left(6-\xi^2\right)}},\quad \nu=0,\\
~\\
\displaystyle{\frac{2\nu\left(\nu+1\right)-\xi^2}{2\left(\nu+1\right)\xi}},\quad \nu\ge1.
\end{array}
\right.
\label{eq:ratapprox}
\eeq
Equations (\ref{eq:epsi1}) are obtained by substituting (\ref{eq:ratapprox}) in the dispersion equation (\ref{eq:SS1}), and solving with respect to $\varepsilon_i$ the arising {\em linear} equations.

The Hankel logarithmic derivative in (\ref{eq:Gn}) admits the following small-argument approximation \cite[Eq. (3.24)]{Alpert:2000re}
\beq
G_\nu\left(\xi\right)\sim
\left\{
\begin{array}{lllll}
\left[\log\left(\displaystyle{\frac{-i\xi}{2}}\right)+\gamma\right]^{-1}\xi^{-1}+{\cal O}\left(\xi\right),\quad\nu=0,\\
~\\
-\nu\xi^{-1}+{\cal O}\left(\xi\log\xi\right),\quad \nu=1,\\
~\\
-\nu\xi^{-1}+{\cal O}\left(\xi\right),\quad \nu>1,
\end{array}
\right.
\label{eq:asGn}
\eeq
with ${\cal O}$ denoting the Landau symbol.
The approximations in (\ref{eq:epsi2}) follow by substituting (\ref{eq:asGn}) in (\ref{eq:epsi1}), and neglecting the higher-order terms.

%%%%%%%%%%%%%%%%%%%%%%%%%%%%%%%%%%%%%%%%%%%%%%%%%%%%%%%%%%%%%%
\section{Details on the Syntheses in Sec. \ref{Sec:RR}}
%%%%%%%%%%%%%%%%%%%%%%%%%%%%%%%%%%%%%%%%%%%%%%%%%%%%%%%%%%%%%%
\label{Sec:AppB}
The basic principle underlying our syntheses in Sec. \ref{Sec:RR} is to numerically optimize the geometrical and constitutive properties so as to maximize selected scattering coefficients and possibly enforce some conditions in the scattering pattern. 

For instance, in the example of Fig. \ref{Figure6}, we assumed the core and shell regions made of gain and loss materials, respectively, and minimized the cost function
\beq
J_c\left({\underline\beta}\right)=\sum_{m=0}^2\left|c_m\right|^{-2},
\label{eq:PS}
\eeq
where ${\underline \beta}=\left[R_i,R_e,\varepsilon_e,\varepsilon'_i,\varepsilon''_{i,0},\varepsilon''_{i,1},\varepsilon''_{i,2}\right]$ is the parameter vector, with each of the scattering coefficients $c_m$ computed by assuming a different gain level $\varepsilon''_{i,m}$. For the geometrical and constitutive parameters, we assumed the following constraints: $0.1\lambda<R_i<3\lambda$,
$0.5\lambda<R_e<3\lambda$
(with $R_i<R_e$), $0.1<\varepsilon'_e<6$, $0.01<\varepsilon''_e<2$,
$0.1<\varepsilon'_i<5$, $-1.1<\varepsilon''_i<-0.01$.

For the example in Fig. \ref{Figure7} (zero forward scattering), we minimized the cost function 
\beq
J_c\left({\underline\beta}\right)=\alpha_0 W_s\left(0\right)
+ \alpha_1\left(\left|c_0\right|^{-2}+\left|c_1\right|^{-2}\right) + \alpha_2\sum_{m=2}^6\left|c_m\right|^2,
\label{eq:ZFS}
\eeq
with the parameter vector ${\underline \beta}=\left[R_i,R_e,\varepsilon_i,\varepsilon_e\right]$ and weight coefficients
$\alpha_0=1$, $\alpha_1=0.05$, $\alpha_2=0.01$. In this case the idea is to minimize the forward scattering while ensuring the dominant character of the two lowest multipolar orders. 
In this and the following examples, we assumed the core and shell regions made of loss and gain materials, respectively, as this configuration was generally found less prone to instability. 
Moreover, we assumed throughout the following constraints for the constitutive parameters: $1<\varepsilon'_i<12$, $0.1<\varepsilon''_i<3$,
$0.1<\varepsilon'_e<5$, $-2<\varepsilon''_i<-0.01$.
Finally, we also enforced some constraints in the strength of the scattering coefficients ($|c_m|\le 10$) so as to sufficiently detune the spectral singularities associated with the dominant orders, and thus favor stability (see the discussion in Sec. \ref{Sec:Stability}).

Along the same lines, for the example in Fig. \ref{Figure8} (transverse scattering), we minimized the cost function
\begin{eqnarray}
J_c\left({\underline\beta}\right) &=& \alpha_0\left[W_s\left(0\right)+W_s\left(\pi\right)\right] + \frac{\alpha_1}{W_s\left(\frac{\pi}{2}\right)}\nonumber\\
&\!\!\!+\!&\!\!\!\!\!\!\alpha_2\!\left(\left|c_0\right|^{-2}\!\!+\!\left|c_2\right|^{-2}\right)\!\!+\!\alpha_3\!\left(\!\left|c_1\right|^2\!\!+\!\!\sum_{m=3}^6\!\left|c_m\right|^2\!\right)\!\!,
\end{eqnarray}
with weight coefficients $\alpha_0=\alpha_1=1$, $\alpha_2=0.05$, $\alpha_3=0.01$. The only differences with respect to (\ref{eq:ZFS}) are the presence of a further condition to penalize the backward scattering and the dominance of the $m=0,2$ multipolar orders.

Finally, for the example in Fig. \ref{Figure9} (pattern reconfigurability), 
\beq
J_c\left({\underline\beta}\right)=W_{s0}\left(0\right)+W_{s1}\left(\frac{\pi}{6}\right)+W_{s2}\left(\frac{\pi}{3}\right),
\label{eq:ZT}
\eeq
with the parameter vector ${\underline \beta}=\left[R_i,R_e,\varepsilon_i,\varepsilon'_e,\varepsilon''_{e,0},\varepsilon''_{e,1},\varepsilon''_{e,2}\right]$, with each of the scattering widths $W_{s\nu}$ computed by assuming a different gain level $\varepsilon''_{e,\nu}$. 

For the minimization of the cost functions in (\ref{eq:PS})--(\ref{eq:ZT}), we followed the approach already successfully utilized in \cite{Moccia:2016de}, relying on the \texttt{NMinimize} function in Mathematica \cite{Mathematica}, which implements a combination of the Nelder-Meald (simplex) and differential-evolution unconstrained optimization algorithms. To cope with the inherent  nonlinear character of the minimization problem, we implemented a synthesis strategy that explores different regions of the search-space via
randomly re-initialization the initial guess across a reasonable parameter range. Accordingly, we enforced the parameter constraints in a soft fashion, via suitable choices of the initial-guess parameter ranges, with {\em a posteriori} verification.

% or


\ifCLASSOPTIONcaptionsoff
  \newpage
\fi



%\bibliographystyle{IEEEtran}
%\bibliography{Moccia_CYL-NH.bib}
%\begin{thebibliography}{1}


\end{document}